\documentclass[aip,reprint]{revtex4-1}
\usepackage[T1]{fontenc}
\usepackage[utf8x]{inputenc}
\usepackage{amsmath} 
\usepackage{graphicx}
\usepackage{siunitx}
\usepackage{xcolor}
\usepackage{amssymb}
\usepackage{amsfonts}
\usepackage{relsize}
\usepackage{mathrsfs}
\usepackage{microtype}
\usepackage[pdfpagemode=UseNone,colorlinks=true,linkcolor=blue,citecolor=blue]{hyperref}
\usepackage{geometry}
\usepackage{eso-pic}
\usepackage{lipsum}
\usepackage{ulem}

\DeclareSIUnit[] \dBm{\rm{dBm}}

\draft

\begin{document}

\title{Mechanical characterization of a membrane with an on-chip loss shield in a cryogenic environment}

\author{Francesco Marzioni}
\affiliation{Physics Division, School of Science and Technology, University of Camerino, I-62032 Camerino, MC, Italy}
\affiliation{INFN, Sezione di Perugia, via A. Pascoli, I-06123 Perugia, Italy}
\affiliation{Department of Physics, University of Naples “Federico II”, I-80126 Napoli, Italy}

\author{Riccardo Natali}
\affiliation{Physics Division, School of Science and Technology, University of Camerino, I-62032 Camerino, MC, Italy}
\affiliation{INFN, Sezione di Perugia, via A. Pascoli, I-06123 Perugia, Italy}

\author{Michele Bonaldi}
\affiliation{Institute of Materials for Electronics and Magnetism, Nanoscience-Trento-FBK Division, 38123 Povo, TN, Italy}
\affiliation{Istituto Nazionale di Fisica Nucleare, TIFPA, 38123 Povo, TN, Italy}

\author{Antonio Borrielli}
\affiliation{Institute of Materials for Electronics and Magnetism, Nanoscience-Trento-FBK Division, 38123 Povo, TN, Italy}
\affiliation{Istituto Nazionale di Fisica Nucleare, TIFPA, 38123 Povo, TN, Italy}

\author{Enrico Serra}
\affiliation{Istituto Nazionale di Fisica Nucleare, TIFPA, 38123 Povo, TN, Italy}
\affiliation{Microelectronics Department, EEMCS Faculty, Delft University of Technology,  2628 CD Delft, The Netherlands}

\author{Bruno Morana}
\affiliation{Microelectronics Department, EEMCS Faculty, Delft University of Technology,  2628 CD Delft, The Netherlands}

\author{Francesco Marin}
\affiliation{CNR-INO, L.go Enrico Fermi 6, I-50125 Firenze, FI, Italy}
\affiliation{INFN, Sezione di Firenze, Via Sansone 1, I-50019 Sesto Fiorentino, Italy}
\affiliation{Dipartimento di Fisica e Astronomia, Università di Firenze, Via Sansone 1, I-50019 Sesto Fiorentino, Italy}

\author{Francesco Marino}
\affiliation{CNR-INO, L.go Enrico Fermi 6, I-50125 Firenze, FI, Italy}
\affiliation{INFN, Sezione di Firenze, Via Sansone 1, I-50019 Sesto Fiorentino, Italy}

\author{Nicola Malossi}
\affiliation{Physics Division, School of Science and Technology, University of Camerino, I-62032 Camerino, MC, Italy}
\affiliation{INFN, Sezione di Perugia, via A. Pascoli, I-06123 Perugia, Italy}

\author{David Vitali}
\affiliation{Physics Division, School of Science and Technology, University of Camerino, I-62032 Camerino, MC, Italy}
\affiliation{INFN, Sezione di Perugia, via A. Pascoli, I-06123 Perugia, Italy}
\affiliation{CNR-INO, L.go Enrico Fermi 6, I-50125 Firenze, FI, Italy}

\author{Giovanni Di Giuseppe}
\affiliation{Physics Division, School of Science and Technology, University of Camerino, I-62032 Camerino, MC, Italy}
\affiliation{INFN, Sezione di Perugia, via A. Pascoli, I-06123 Perugia, Italy}

\author{Paolo Piergentili}
\email{paolo.piergentili@unicam.it}
\affiliation{Physics Division, School of Science and Technology, University of Camerino, I-62032 Camerino, MC, Italy}
\affiliation{INFN, Sezione di Perugia, via A. Pascoli, I-06123 Perugia, Italy}

\date{\today}

\begin{abstract}
The quantum transduction of an rf/microwave signal to the optical domain, and vice versa, paves the way for technologies that exploit the advantages of each domain to perform quantum operations.
Since electro-optomechanical devices implement a simultaneous coupling of a mechanical oscillator to both an rf/microwave field and an optical field, they are suitable for the realization of a quantum transducer.
The membrane-in-the-middle setup is a possible solution, once its vibrational mode is cooled down to ultra cryogenic temperature for achieving quantum operation.
This work is focused on the mechanical characterization via an optical interferometric probe, down to $T=\SI{18}{\milli\kelvin}$, of a loss-shielded metalized membrane designed for this purpose.
A stroboscopic technique has been exploited for revealing a mechanical quality factor up to $\num{64e6}$ at the lowest temperature.
In fact, with continuous illumination and a cryostat temperature below $\SI{1}{\kelvin}$, the heat due to optical absorption is not efficiently dissipated anymore, and the membrane remains hotter than its environment. 
\end{abstract}


\maketitle

Quantum transduction refers to the process of converting one form of energy to another at the single excitation level, and it represents a key ingredient in quantum technologies.
Quantum transduction could be used for the optical detection of microwave/rf signals by exploiting the most efficient detectors for optical photons.\cite{Regal_2011,PhysRevLett.107.273601,Bagci:2014aa,Takeda:18,PhysRevApplied.9.034031,Simonsen:19,Simonsen:2019aa}
Moreover, a major interest is currently placed on the coherent conversion between optical and microwave/radio frequency (mw/rf) photons,\cite{PhysRevA.84.042342,10.1063/5.0021088,Lauk_2020} the optical domain being ideal for reliable long-range communications through optical fibers or in free space, while the lower frequency band is particularly suitable for high-fidelity local quantum operations using superconducting and other solid-state processors.
This will allow a global quantum Internet or distributed quantum tasks including computing or sensing.\cite{Kimble:2008aa,Pirandola:2018aa,app14010387}
The easiest way to bridge the enormous energy gap is to use a mediator simultaneously coupled to both microwave/rf and optical modes. 
There has been a variety of proposals using different kinds of mediating systems.
Here, we focus on electro-optomechanical systems,\cite{PhysRevA.84.042342,PhysRevLett.108.153603,PhysRevLett.108.153604,PhysRevLett.109.130503,Hill:2012aa,Andrews:2014aa,Higginbotham:2018aa,Planz:22} where a mechanical resonator is capacitively coupled to mw/rf photons,\cite{PhysRevLett.107.273601,Bagci:2014aa,Takeda:18,PhysRevApplied.9.034031,Simonsen:19,Simonsen:2019aa,Andrews:2014aa,Higginbotham:2018aa,PhysRevA.103.033516,e25071087,PhysRevX.12.021062} and dispersively via radiation pressure with the optical mode(s).
Various solutions have been adopted for the explicit design of electro-optomechanical transducers.
One approach makes use of direct electromechanical capacitive coupling in metalized membranes \cite{PhysRevLett.107.273601,Bagci:2014aa,Takeda:18,PhysRevApplied.9.034031,Simonsen:19,Simonsen:2019aa,PhysRevLett.107.273601,Bagci:2014aa,Takeda:18,PhysRevApplied.9.034031,Simonsen:19,Simonsen:2019aa,e25071087,PhysRevX.12.021062,Seis:2022aa} which, with an in-front electrode, are placed within an optical Fabry--Pérot cavity for radiation pressure coupling.
This membrane-in-the-middle configuration has the potential to exploit cooperative and coupling enhancement effects, which can be obtained when two parallel membrane resonators are placed within a Fabry--Pérot cavity.\cite{Piergentili_2018,Gartner:2018aa,PhysRevLett.124.053604,PhysRevApplied.15.034012,Piergentili_2021,deJong:22,Piergentili:2022aa,10.3389/fphy.2023.1222056,PhysRevA.111.013525,PhysRevX.15.011014}

The heating of the mechanical system due to the absorption of electromagnetic radiation can be detrimental, especially in the optical domain, where the energy of the photons is higher with respect to microwaves.
A further limitation may appear when working with patterned mechanical resonators, which can show higher thermal isolation than plain surfaces.\cite{10.1063/1.120979,Planz:23}
This effect has been recently studied in several optomechanical and electromechanical platforms, such as optomechanical crystals,\cite{doi:10.1126/science.abc7312} whispering gallery mode resonators,\cite{PRXQuantum.1.020315} in-the-middle micromechanical silicon nitride membranes,\cite{Zwickl:2008aa,PhysRevLett.116.063601,Planz:23} micropillars,\cite{10.1063/1.4863666} beam modes\cite{PhysRevApplied.12.044066} and so on.

In this work we will present the mechanical characterization with an optical interferometric probe, from room temperature down to $\SI{18}{\milli\kelvin}$, of a loss-shielded silicon nitride (Si$_3$N$_4$) membrane, which is partially covered with a titanium nitride (TiNx) overlay.\cite{e25071087}
The metal layer is designed for realizing a capacitor with an in-front electrode, and at the same time to allow for a window of silicon nitride at the center of the membrane.
This solution allows the simultaneous coupling of the membrane both to an rf/microwave resonator via capacitive coupling and to an optical cavity via radiation pressure.
Large electromechanical and optomechanical cooperativities are required to operate the electro-optomechanical device in the quantum regime,\cite{PhysRevA.103.033516,RevModPhys.86.1391} and the mechanical quality factor, being proportional to both, plays a key role.
Here, we describe a detailed experimental analysis of the mechanical quality factor, showing that it increases around 10 times when the temperature drops from $\SI{1}{\kelvin}$ to $\SI{20}{\milli\kelvin}$, provided that stroboscopic light is employed to measure the motion of the membrane.\cite{Rossi:2018aa, Liu:2025aa}
Our results reveal the dependence of the mechanical quality factor of the device from room temperature down to the ultracryogenics range.
Below $\SI{1}{\kelvin}$, the effect of the infrared radiation directly impinging on the membrane is observed.
Finally, we note that our results show a feature of thermoelastic damping in the silicon support frame, indicating that it is the dominant loss in the range of temperatures within $\SI{25}{\kelvin}$ and $\SI{175}{\kelvin}$.\cite{10.1063/1.2868810} 

\begin{figure}[h]
\includegraphics[width=.45\textwidth]{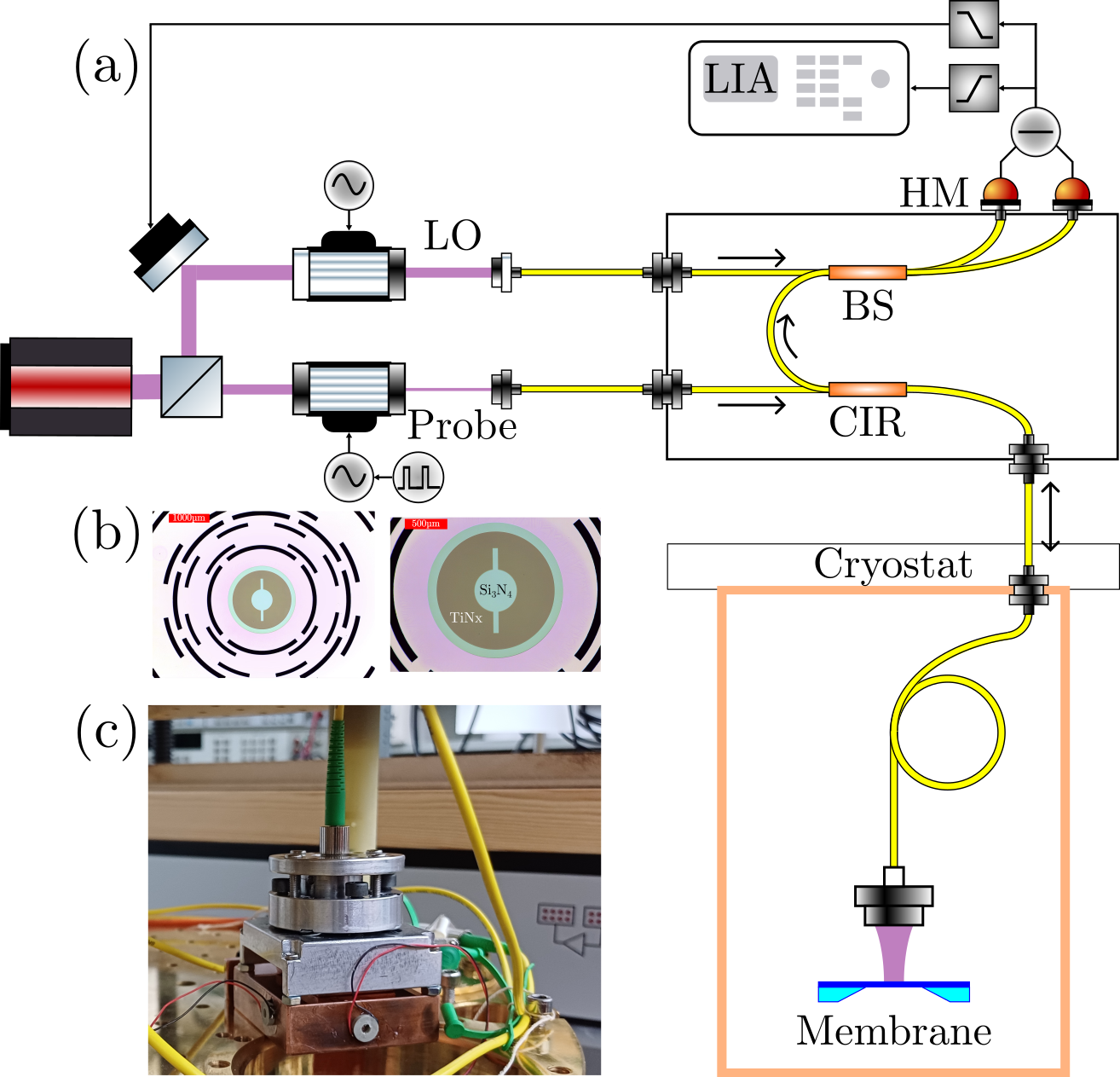}
\caption{(a) Sketch of the experimental setup. A Mach-Zehnder interferometer (BS denotes a beam splitter, CIR a circulator) detects the mechanical oscillations of the membrane inside a dilution cryostat's mixing chamber. The local oscillator (LO) and the probe (Probe) beams of the interferometer are modulated independently by two acousto-optic modulators (AOMs) driven at the same frequency and in phase. The intensity of the probe beam can be constant for an uninterrupted interferometric detection of the membrane motion or modulated by the train of pulses that drives the tone generator if a stroboscopic illumination of the sample is required. The balanced homodyne (HM) detection signal is low-pass filtered for phase-locking the interferometer and high-pass filtered for the lock-in amplifier (LIA) analysis. (b) On the left, a schematic of the Si$_3$N$_4$ membrane (light blue) with the TiNx overlay (brown) is illustrated. The membrane is endowed with the on-chip shield. On the right, a detailed view of the TiNx layer is shown. Two red rulers are placed on the top left corners of the two panels, and they indicate $\SI{1}{\milli\meter}$ and $\SI{0.5}{\milli\meter}$ on the left one and on the right one, respectively. (c) Image of the membrane in its support, attached on the cryostat plate, with the optical fiber connected.}
\label{fig:interf}
\end{figure}

The experimental setup consists of a Mach--Zehnder interferometer, presented schematically in Fig.~\ref{fig:interf}(a).
A local oscillator (LO) and a probe beam (Probe) are separated from the same $\SI{1064}{\nano\meter}$ laser in free space, modulated by means of two acousto-optic-modulators (AOMs) driven by two external tones from an arbitrary function generator at the same frequency and phase.
The generation of a train of pulses on the probe, to perform stroboscopic measurements, is controlled by cascade-driving the tone generator with a square shaped waveform.
After the modulation, the two beams are fiber coupled.
The probe is sent to a circulator (CIR) port, and the output of the CIR to the cryostat for reading out the motion of a metalized membrane.
As shown in Fig.~\ref{fig:interf}(b), the membrane is a $\SI{80\pm5}{\nano\meter}$ thick silicon nitride (Si$_3$N$_4$) layer, which is partially covered with a titanium nitride (TiNx) overlay deposited with a target thickness of $\SI{50}{\nano\meter}$,\cite{e25071087} giving an effective mass of the fundamental mode $m_{\mathrm{eff}}=\SI{2.7e-10}{\kilo\gram}$.
The surface tension of the membrane is around $\SI{1}{\giga\pascal}$, and we find the first resonance peak at $\SI{265}{\kilo\hertz}$.
The mechanical dissipation is reduced using an on-chip shield, which insulates the membrane from the substrate.\cite{10.1063/5.0055954}
The membrane is mounted inside the mixing chamber of the dilution refrigerator, with its frame thermally connected to the cryostat's plate, as illustrated in Fig.~\ref{fig:interf}(c).
The probe beam is focused to the center of the membrane using a lens fiber coupler (Thorlabs CFC2).
The reflected light is selected by the CIR and sent to one port of a fiber beam splitter (BS).
The LO enters from the other port of the BS, and the interference at its outputs is detected.
The differential detection of the two photocurrents results in the homodyne (HM) signal.
The low-frequency component of the signal is filtered, amplified, and fed back to stabilize the interferometer. 
The high-frequency range, instead, is analyzed by means of a lock-in amplifier (LIA).

\begin{figure}[h]
\centering
\includegraphics[width=.45\textwidth]{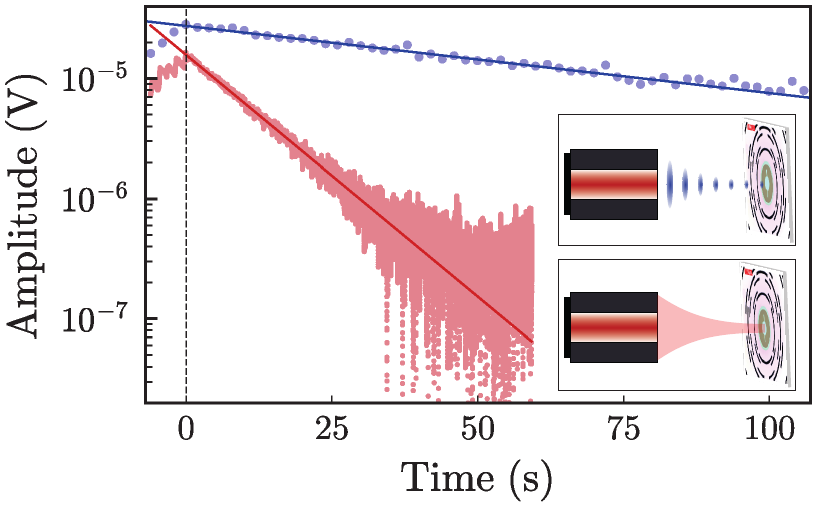}
\caption{Typical ring-down measurement at $T=\SI{20}{\milli\kelvin}$. The red points show the amplitude decay of the membrane motion when the membrane is continuously illuminated by the probe light. The light blue points represent the ring-down when the membrane motion is stroboscopically detected. The two cases are illustrated in the inset. The best-fitting lines are illustrated in dark red and blue, respectively. The ring-down starting point corresponds to the zero of the time axis and it is marked by the black-dashed vertical line.}
\label{fig:typicalmeas}
\end{figure}

\begin{figure}[htbp]
\centering
\includegraphics[width=.45\textwidth]{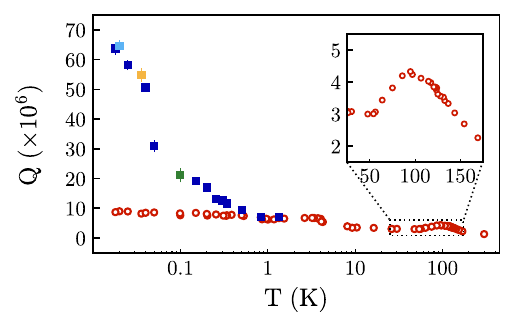}
\caption{Measured mechanical quality factor $\mathrm{Q}$ of the metalized membrane as a function of the temperature of the cryostat plate $T$. The red empty circles show the quality factor measured by continuously reading out the motion of the membrane, while the blue squares stand for stroboscopic readout. The data of the stroboscopic measurements at $T=\SI{20}{\milli\kelvin}$, $T=\SI{35}{\milli\kelvin}$, and $T=\SI{99}{\milli\kelvin}$ are highlighted with different colors, related to the results shown in Fig.~\ref{fig:Qfactor_dc}. The uninterrupted beam power, which corresponds to the peak power of the stroboscopic pulses, is $P\simeq\SI{100}{\nano\watt}$. The inset highlights the data that fall in the range of temperatures within $\SI{25}{\kelvin}$ and $\SI{175}{\kelvin}$, showing a feature of thermoelastic damping in the silicon support frame.\cite{10.1063/1.2868810}}
\label{fig:Qfactor_Tlog}
\end{figure}

We studied the mechanical features of the metalized membrane as a function of the temperature of the cryostat plate. 
We performed the mechanical characterization over a wide range of temperature, from room temperature down to $\SI{18}{\milli\kelvin}$, by measuring the mechanical quality factor $\mathrm{Q}$ of the membrane.
The quality factors are estimated by measuring the ring-down time of the membrane's fundamental mode.
A typical measurement dataset, with the cryostat plate at $\SI{20}{\milli\kelvin}$, is shown in Fig.~\ref{fig:typicalmeas}.
In our setup, a piezoelectric shaker actuates the motion of the fundamental mode of the membrane in the linear regime.
Once the excitation tone has been switched off, we measure the free evolution of the amplitude of the motion by means of a lock-in amplifier (Zurich Instruments HF2LI), which detects and filters the homodyne signal.
We determine the characteristic decay time of the mechanical motion, or ring-down time, $\tau_{\rm{rd}}$, by fitting the data with an exponential function.
The quality factor of the oscillator is evaluated as $\rm{Q}=\pi\nu\tau_{\rm{rd}}$, where $\nu$ is the resonance frequency of the mechanical mode.
In Fig.~\ref{fig:Qfactor_Tlog}, the measured $\rm{Q}$-factor as a function of the temperature of the cryostat plate is shown.
To thermalize the membrane with the cryostat plate and obtain the vacuum inside the cryostat, the measurements are taken few hours after we changed the cryostat temperature.
Also, the probe beam is turned on few minutes before each measurement.
The piezoelectric actuator and the light are both heat sources for the cryostat.
However, we have verified that at $P\simeq\SI{100}{\nano\watt}$ the light scattered in the fridge does not heat the mixing chamber plate, and we have driven the piezoelectric with short pulses to reduce the power dissipated on the plate. 
During the measurements we do not observe any significant temperature variation of the cryostat's plate.

The continuous interferometric detection of the membrane motion via homodyne detection implies the uninterrupted illumination of the membrane, and it represents a detrimental effect in the range of temperatures roughly below $\SI{1}{\kelvin}$.
In fact, as clearly pointed out by the red empty circles in Fig.~\ref{fig:Qfactor_Tlog}, the measured quality factor of the continuously illuminated membrane flattens around $\rm{Q}=\num{8e6}$ for temperatures below $\SI{1}{\kelvin}$.
However, if a stroboscopic technique is employed to detect the mechanical motion,\cite{Rossi:2018aa} the ring-down is longer than the one measured with continuous light, as shown by the blue points with respect to the red points in Fig.~\ref{fig:typicalmeas}.
The measured quality factor is noticeably higher with the stroboscopic method, reaching $\rm{Q}_{\rm{max}}=\num{64\pm2e6}$ at $T=\SI{18}{\milli\kelvin}$, as illustrated by the blue squares in Fig.~\ref{fig:Qfactor_Tlog}.
The stroboscopic technique consists of illuminating the membrane with a train of pulses of duration $\tau_{\rm{s}}$ and period $\rm{T}_{\rm{s}}$.
In this experiment we have estimated the quality factors at $T=\SI{18}{\milli\kelvin}$, $\SI{20}{\milli\kelvin}$, $\SI{25}{\milli\kelvin}$, $\SI{35}{\milli\kelvin}$, $\SI{40}{\milli\kelvin}$, $\SI{50}{\milli\kelvin}$, and $\SI{99}{\milli\kelvin}$ using pulses of duration $\tau_{\rm{s}}=\SI{40}{\milli\second}$ and period $\rm{T}_{\rm{s}}=\SI{2}{\second}$.
The other stroboscopic data points are obtained with pulses of duration $\tau_{\rm{s}}=\SI{100}{\milli\second}$ and period $\rm{T}_{\rm{s}}=\SI{2}{\second}$.
These parameters permit a good reconstruction of the mechanical amplitude decay, and at the same time an effective measurement of the quality factors.

\begin{figure}[h]
\centering
\includegraphics[width=.45\textwidth]{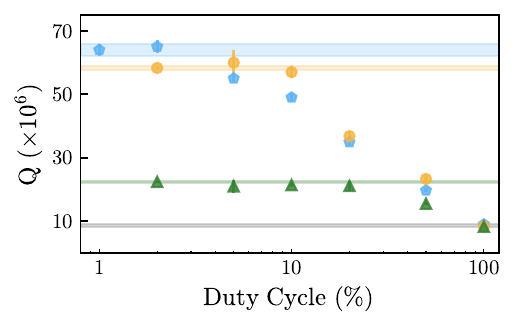}
\caption{Measured mechanical quality factor $\mathrm{Q}$ of the metalized membrane as a function of the duty cycle at different temperatures. Light blue pentagons, orange circles, and green triangles indicate the data measured with the cryostat plate at $\SI{20}{\milli\kelvin}$, $\SI{35}{\milli\kelvin}$, and $\SI{99}{\milli\kelvin}$, respectively. The shaded horizontal lines, color associated with the data, illustrate the value of the quality factor used in Fig.~\ref{fig:Qfactor_Tlog} for the respective temperature. The gray horizontal line shows the mechanical quality factor for continuous illumination with $\SI{100}{\nano\watt}$ light power, which coincides for all the temperatures below $\SI{1}{\kelvin}$.}
\label{fig:Qfactor_dc}
\end{figure}

The behavior of the measured mechanical quality factor vs the plate temperature illustrated in Fig.~\ref{fig:Qfactor_Tlog} highlights the detrimental effects of the continuous illumination for temperatures below $\SI{1}{\kelvin}$.
Our explanation is that, below this temperature, the thermal conductivity of the device drops to very low values and the optical power absorbed by the membrane is not efficiently dissipated anymore. 
As a consequence, the membrane is heated, and the local membrane temperature remains much higher than that of the fridge plate.
We have studied the effect of the duration of the pulses, keeping the period $\rm{T}_{\rm{s}}=\SI{2}{\second}$ fixed.
We indicate the duration of the pulses as the duty cycle (dc), which is the percentage given by $\rm{dc}=(\tau_{\rm{s}}/\rm{T}_{\rm{s}})\times100\%$.
The experimental data for $T=\SI{20}{\milli\kelvin}$, $\SI{35}{\milli\kelvin}$, and $\SI{99}{\milli\kelvin}$ are reported in Fig.~\ref{fig:Qfactor_dc} as light blue pentagons, orange circles, and green triangles, respectively.
The measured behavior confirms our explanation in terms of a much higher effective membrane temperature.
In principle, lower light powers would be less invasive, also employing an uninterrupted detection, but we are limited by the locking error signal of the homodyne detector, whose amplitude lowers as the light shining the membrane decreases. 
For our experimental setup the optimal power is $\SI{100}{\nano\watt}$.

The lack of a cavity, which properly enhances the signal, does not allow the thermal motion of the oscillator to be revealed, and this prevents the estimation of the effective temperature of the membrane mode by previous knowledge of its mass.

In conclusion, we reported the mechanical characterization of a metalized membrane attached to the plate of a dilution refrigerator.
Our results showed that the light of the interferometer probe beam influences the measured quality factor for temperatures of the cryostat plate below $T=\SI{1}{\kelvin}$.
However, making use of stroboscopic light, we were able to measure a mechanical quality factor of $\num{64e6}$ at $T=\SI{18}{\milli\kelvin}$, which is almost 10 times higher than the one measured with continuous light.
The proposed source for this effect is the heating of the membrane due to the light: since the thermal conductivities of the materials drop at ultracryogenic temperatures, even with low light absorption, the temperature of the membrane significantly increases.
We presented an analysis of the dependence of the mechanical quality factor on the duration of the pulses of the stroboscopic illumination, which supports our explanation.
The stroboscopic method employed here provides results consistent with those obtained with microwave excitation,\cite{10.1063/1.4938747} and it shows the need of a noninvasive measurement when adopting optical readout in an ultracryogenic domain (see also Refs.~\onlinecite{Planz:23} and~\onlinecite{doi:10.1126/science.abc7312}).
This characterization is an experimental key point for the realization of an electro-optomechanical quantum transducer between the optical and the rf domains, which exploits the membrane-in-the-middle dispersive coupling to a high-finesse optical cavity and the capacitive coupling to an LC circuit.

\begin{acknowledgments}
We acknowledge the support of the PNRR MUR Project No. PE0000023-NQSTI (Italy).
\end{acknowledgments}

\section*{Author Declarations}
\subsection*{Conflict of Interest}
The authors have no conflicts to disclose.
\subsection*{Author Contributions}
\textbf{Francesco Marzioni:} Data curation (equal); Formal analysis (equal); Investigation (equal); Methodology (equal); Validation (equal); Visualization (equal); Writing - original draft (equal); Writing - review \& editing (equal).
\textbf{Riccardo Natali:} Investigation (equal); Methodology (equal); Writing - review \& editing (equal).
\textbf{Michele Bonaldi:} Resources (equal); Writing - review \& editing (equal).
\textbf{Antonio Borrielli:} Resources (equal); Writing - review \& editing (equal).
\textbf{Enrico Serra:} Resources (equal); Writing - review \& editing (equal).
\textbf{Bruno Morana:} Resources (equal); Writing - review \& editing (equal).
\textbf{Francesco Marin:} Funding acquisition (equal); Project administration (equal); Writing - review \& editing (equal).
\textbf{Francesco Marino:} Writing - review \& editing (equal).
\textbf{Nicola Malossi:} Conceptualization (equal); Investigation (equal); Methodology (equal); Writing - review \& editing (equal).
\textbf{David Vitali:} Conceptualization (equal); Funding acquisition (equal); Project administration (equal); Writing - review \& editing (equal).
\textbf{Giovanni Di Giuseppe:} Conceptualization (equal); Investigation (equal); Methodology (equal); Supervision (equal); Validation (equal); Writing - original draft (equal); Writing - review \& editing (equal).
\textbf{Paolo Piergentili:} Conceptualization (equal); Data curation (equal); Formal analysis (equal); Investigation (equal); Methodology (equal); Supervision (equal); Validation (equal); Writing - original draft (equal); Writing - review \& editing (equal).
\section*{Data Availability}
The data that support the findings of this study are available from the corresponding author upon reasonable request.

\section*{References}

\begin{thebibliography}{10}

\bibitem{Regal_2011}
C~A Regal and K~W Lehnert,
\newblock ``From cavity electromechanics to cavity optomechanics'',
\newblock {\em J.~Phys.~Conf.~Ser.} 264, 012025 (2011).

\bibitem{PhysRevLett.107.273601}
J.~M. Taylor, A.~S. S\o{}rensen, C.~M. Marcus, and E.~S. Polzik,
\newblock ``Laser cooling and optical detection of excitations in a $lc$ electrical circuit'',
\newblock {\em Phys.~Rev.~Lett.} 107, 273601 (2011).

\bibitem{Bagci:2014aa}
T.~Bagci, A.~Simonsen, S.~Schmid, L.~G.~Villanueva, E.~Zeuthen, J.~Appel, J.~M.~Taylor, A.~S{\o}rensen, K.~Usami, A.~Schliesser, and E.~S.~Polzik,
\newblock ``Optical detection of radio waves through a nanomechanical transducer'',
\newblock {\em Nature} 507, 81--85 (2014).

\bibitem{Takeda:18}
K.~Takeda, K.~Nagasaka, A.~Noguchi, R.~Yamazaki, Y.~Nakamura, E.~Iwase, J.~M.~Taylor, and K.~Usami,
\newblock ``Electro-mechano-optical detection of nuclear magnetic resonance'',
\newblock {\em Optica} 5, 152--158 (2018).

\bibitem{PhysRevApplied.9.034031}
I.~Moaddel~Haghighi, N.~Malossi, R.~Natali, G.~Di~Giuseppe, and D.~Vitali,
\newblock ``Sensitivity-bandwidth limit in a multimode optoelectromechanical transducer'',
\newblock {\em Phys.~Rev.~Appl.} 9, 034031 (2018).

\bibitem{Simonsen:19}
A.~Simonsen, S.~A.~Saarinen, J.~D.~Sanchez, J.~H.~Ardenkj{\ae}r-Larsen, A.~Schliesser, and E.~S.~Polzik,
\newblock ``Sensitive optomechanical transduction of electric and magnetic signals to the optical domain'',
\newblock {\em Opt.~Express} 27, 18561--18578 (2019).

\bibitem{Simonsen:2019aa}
A.~Simonsen, J.~D.~S{\'a}nchez-Heredia, S.~A.~Saarinen, J.~H.~Ardenkj{\ae}r-Larsen, A.~Schliesser, and E.~S. Polzik,
\newblock ``Magnetic resonance imaging with optical preamplification and detection'',
\newblock {\em Sci.~Rep.} 9, 18173 (2019).

\bibitem{PhysRevA.84.042342}
S.~Barzanjeh, D.~Vitali, P.~Tombesi, and G.~J.~Milburn,
\newblock ``Entangling optical and microwave cavity modes by means of a nanomechanical resonator'',
\newblock {\em Phys.~Rev. A} 84, 042342 (2011).

\bibitem{10.1063/5.0021088}
Y.~Chu and S.~Gr{\"o}blacher,
\newblock ``A perspective on hybrid quantum opto- and electromechanical systems'',
\newblock {\em App.~Phys.~Lett.} 117, 150503 (2020).

\bibitem{Lauk_2020}
N.~Lauk, N.~Sinclair, S.~Barzanjeh, J.~P.~Covey, M.~Saffman, M.~Spiropulu, and C.~Simon,
\newblock ``Perspectives on quantum transduction'',
\newblock {\em Quantum~Sci.~Technol.} 5, 020501 (2020).

\bibitem{Kimble:2008aa}
H.~J.~Kimble,
\newblock ``The quantum internet'',
\newblock {\em Nature} 453, 1023--1030 (2008).

\bibitem{Pirandola:2018aa}
S.~Pirandola, B.~R.~Bardhan, T.~Gehring, C.~Weedbrook, and S.~Lloyd,
\newblock ``Advances in photonic quantum sensing'',
\newblock {\em Nat.~Photonics} 12, 724--733 (2018).

\bibitem{app14010387}
P.~Piergentili, F.~Amanti, G.~Andrini, F.~Armani, V.~Bellani, V.~Bonaiuto, S.~Cammarata, M.~Campostrini, S.~Cornia, T.~H.~Dao, F.~De~Matteis, V.~Demontis, G.~Di~Giuseppe, S.~Ditalia~Tchernij, S.~Donati, A.~Fontana, J.~Forneris, R.~Francini, L.~Frontini, R.~Gunnella, S.~Iadanza, A.~E.~Kaplan, C.~Lacava, V.~Liberali, F.~Marzioni, E.~Nieto~Hern{\'a}ndez, E.~Pedreschi, D.~Prete, P.~Prosposito, V.~Rigato, C.~Roncolato, F.~Rossella, A.~Salamon, M.~Salvato, F.~Sargeni, J.~Shojaii, F.~Spinella, A.~Stabile, A.~Toncelli, G.~Trucco, and V.~Vitali,
\newblock ``Quantum information with integrated photonics'',
\newblock {\em Appl.~Sci.} 14, 387 (2024).

\bibitem{PhysRevLett.108.153603}
Y.~D.~Wang and A.~A.~Clerk,
\newblock ``Using interference for high fidelity quantum state transfer in optomechanics'',
\newblock {\em Phys.~Rev.~Lett.} 108, 153603 (2012).

\bibitem{PhysRevLett.108.153604}
L.~Tian,
\newblock ``Adiabatic state conversion and pulse transmission in optomechanical systems'',
\newblock {\em Phys.~Rev.~Lett.} 108, 153604 (2012).

\bibitem{PhysRevLett.109.130503}
S.~Barzanjeh, M.~Abdi, G.~J.~Milburn, P.~Tombesi, and D.~Vitali,
\newblock ``Reversible optical-to-microwave quantum interface'',
\newblock {\em Phys.~Rev.~Lett.} 109, 130503 (2012).

\bibitem{Hill:2012aa}
J.~T.~Hill, A.~H.~Safavi-Naeini, J.~Chan, and O.~Painter,
\newblock ``Coherent optical wavelength conversion via cavity optomechanics'',
\newblock {\em Nat.~Commun.} 3, 1196 (2012).

\bibitem{Andrews:2014aa}
R.~W.~Andrews, R.~W.~Peterson, T.~P.~Purdy, K.~Cicak, R.~W.~Simmonds, C.~A.~Regal, and K.~W.~Lehnert,
\newblock ``Bidirectional and efficient conversion between microwave and optical light'',
\newblock {\em Nat.~Phys.} 10, 321--326 (2014).

\bibitem{Higginbotham:2018aa}
A.~P.~Higginbotham, P.~S.~Burns, M.~D.~Urmey, R.~W.~Peterson, N.~S.~Kampel, B.~M.~Brubaker, G.~Smith, K.~W.~Lehnert, and C.~A.~Regal,
\newblock ``Harnessing electro-optic correlations in an efficient mechanical converter'',
\newblock {\em Nat.~Phys.} 14, 1038--1042 (2018).

\bibitem{Planz:22}
E.~Planz, Y.~Seis, T.~Capelle, X.~Xi, E.~Langman, and A.~Schliesser,
\newblock ``Towards quantum electro-optic transduction with an embedded 100 ms coherence time quantum memory'',
\newblock in {\em Quantum~2.0~Conference~and~Exhibition} Optica Publishing Group, p. QM4B.7 (2022).

\bibitem{PhysRevA.103.033516}
N.~Malossi, P.~Piergentili, J.~Li, E.~Serra, R.~Natali, G.~Di~Giuseppe, and D.~Vitali,
\newblock ``Sympathetic cooling of a radio-frequency $\mathit{LC}$ circuit to its ground state in an optoelectromechanical system'',
\newblock {\em Phys.~Rev.~A} 103, 033516 (2021).

\bibitem{e25071087}
M.~Bonaldi, A.~Borrielli, G.~Di~Giuseppe, N.~Malossi, B.~Morana, R.~Natali, P.~Piergentili, P.~M.~Sarro, E.~Serra, and D.~Vitali,
\newblock ``Low noise opto-electro-mechanical modulator for rf-to-optical transduction in quantum communications'',
\newblock {\em Entropy} 25, 1087 (2023).

\bibitem{PhysRevX.12.021062}
B.~M.~Brubaker, J.~M.~Kindem, M.~D.~Urmey, S.~Mittal, R.~D.~Delaney, P.~S.~Burns, M.~R.~Vissers, K.~W.~Lehnert, and C.~A.~Regal,
\newblock ``Optomechanical ground-state cooling in a continuous and efficient electro-optic transducer'',
\newblock {\em Phys.~Rev.~X} 12, 021062 (2022).

\bibitem{Seis:2022aa}
Y.~Seis, T.~Capelle, E.~Langman, S.~Saarinen, E~Planz, and A.~Schliesser,
\newblock ``Ground state cooling of an ultracoherent electromechanical system'',
\newblock {\em Nat.~Commun.} 13, 1507 (2022).

\bibitem{Piergentili_2018}
P.~Piergentili, L.~Catalini, M.~Bawaj, S.~Zippilli, N.~Malossi, R.~Natali, D.~Vitali, and G.~Di Giuseppe,
\newblock ``Two-membrane cavity optomechanics'',
\newblock {\em New~J.~Phys.} 20, 083024 (2018).

\bibitem{Gartner:2018aa}
C.~G{\"a}rtner, J.~P.~Moura, W.~Haaxman, R.~A.~Norte, and S.~Gr{\"o}blacher,
\newblock ``Integrated optomechanical arrays of two high reflectivity sin membranes'',
\newblock {\em Nano~Lett.} 18, 7171--7175 (2018).

\bibitem{PhysRevLett.124.053604}
J.~Sheng, X~Wei, C~Yang, and H.~Wu,
\newblock ``Self-organized synchronization of phonon lasers'',
\newblock {\em Phys.~Rev.~Lett.} 124, 053604 (2020).

\bibitem{PhysRevApplied.15.034012}
P.~Piergentili, W.~Li, R.~Natali, D.~Vitali, and G.~Di~Giuseppe,
\newblock ``Absolute determination of the single-photon optomechanical coupling rate via a hopf bifurcation'',
\newblock {\em Phys.~Rev.~Appl.} 15, 034012 (2021).

\bibitem{Piergentili_2021}
P.~Piergentili, W.~Li, R.~Natali, N.~Malossi, D.~Vitali, and G.~Di~Giuseppe,
\newblock ``Two-membrane cavity optomechanics: non-linear dynamics'',
\newblock {\em New~J.~Phys.} 23, 073013 (2021).

\bibitem{deJong:22}
M.~H.~J.~de~Jong, J.~Li, C.~G\"{a}rtner, R.~A.~Norte, and S.~Gr\"{o}blacher,
\newblock ``Coherent mechanical noise cancellation and cooperativity competition in optomechanical arrays'',
\newblock {\em Optica} 9, 170--176 (2022).

\bibitem{Piergentili:2022aa}
P.~Piergentili, R.~Natali, D.~Vitali, and G.~Di~Giuseppe,
\newblock ``Two-membrane cavity optomechanics: Linear and non-linear dynamics'',
\newblock {\em Photonics} 9, 99 (2022).

\bibitem{10.3389/fphy.2023.1222056}
F.~Marzioni, F.~Rasponi, P.~Piergentili, R.~Natali, G.~Di~Giuseppe, and D.~Vitali,
\newblock ``Amplitude and phase noise in two-membrane cavity optomechanics'',
\newblock {\em Front.~Phys.} 11, 1222056 (2023).

\bibitem{PhysRevA.111.013525}
F.~Marzioni, R.~Natali, N.~Malossi, D.~Vitali, G.~Di~Giuseppe, and P.~Piergentili,
\newblock ``Two-membrane etalon'',
\newblock {\em Phys.~Rev.~A} 111, 013525 (2025).

\bibitem{PhysRevX.15.011014}
X.~Yao, M.~H.~J.~de~Jong, J.~Li, and S.~Gr\"oblacher,
\newblock ``Long-range optomechanical interactions in sin membrane arrays'',
\newblock {\em Phys.~Rev.~X} 15, 011014 (2025).

\bibitem{10.1063/1.120979}
M.~M.~Leivo and J.~P.~Pekola,
\newblock ``Thermal characteristics of silicon nitride membranes at sub-kelvin temperatures'',
\newblock {\em Appl.~Phys.~Lett.} 72, 1305--1307, (1998).

\bibitem{Planz:23}
E.~Planz, X.~Xi, T.~Capelle, E.~C.~Langman, and A.~Schliesser,
\newblock ``Membrane-in-the-middle optomechanics with a soft-clamped membrane at millikelvin temperatures'',
\newblock {\em Opt.~Express} 31, 41773--41782 (2023).

\bibitem{doi:10.1126/science.abc7312}
G.~S.~MacCabe, H.~Ren, J.~Luo, J.~D.~Cohen, H.~Zhou, A.~Sipahigil, M.~Mirhosseini, and O.~Painter,
\newblock ``Nano-acoustic resonator with ultralong phonon lifetime'',
\newblock {\em Science} 370, 840--843 (2020).

\bibitem{PRXQuantum.1.020315}
W.~Hease, A.~Rueda, R.~Sahu, M.~Wulf, G.~Arnold, H.~G.~L.~Schwefel, and J.~M.~Fink,
\newblock ``Bidirectional electro-optic wavelength conversion in the quantum ground state'',
\newblock {\em PRX~Quantum} 1, 020315 (2020).

\bibitem{Zwickl:2008aa}
B.~M.~Zwickl, W.~E.~Shanks, A.~M.~Jayich, C.~Yang, A.~C.~Bleszynski~Jayich, J.~D.~Thompson, and J.~G.~E.~Harris,
\newblock ``High quality mechanical and optical properties of commercial silicon nitride membranes'',
\newblock {\em Appl.~Phys.~Lett.} 92, 103125 (2008).

\bibitem{PhysRevLett.116.063601}
R.~W.~Peterson, T.~P.~Purdy, N.~S.~Kampel, R.~W.~Andrews, P.-L.~Yu, K.~W.~Lehnert, and C.~A.~Regal,
\newblock ``Laser cooling of a micromechanical membrane to the quantum backaction limit'',
\newblock {\em Phys.~Rev.~Lett.} 116, 063601 (2016).

\bibitem{10.1063/1.4863666}
A.~G.~Kuhn, J.~Teissier, L.~Neuhaus, S.~Zerkani, E.~van~Brackel, S.~Del{\'e}glise, T.~Briant, P.-F.~Cohadon, A.~Heidmann, C.~Michel, L.~Pinard, V.~Dolique, R.~Flaminio, R.~Ta{\"\i}bi, C.~Chartier, and O.~Le~Traon,
\newblock ``Free-space cavity optomechanics in a cryogenic environment'',
\newblock {\em Appl.~Phys.~Lett.} 104, 044102 (2014).

\bibitem{PhysRevApplied.12.044066}
X.~Zhou, D.~Cattiaux, R.~R.~Gazizulin, A.~Luck, O.~Maillet, T.~Crozes, J.-F.~Motte, O.~Bourgeois, A.~Fefferman, and E.~Collin,
\newblock ``On-chip thermometry for microwave optomechanics implemented in a nuclear demagnetization cryostat'',
\newblock {\em Phys.~Rev.~Appl.} 12, 044066 (2019).

\bibitem{RevModPhys.86.1391}
M.~Aspelmeyer, T.~J.~Kippenberg, and F.~Marquardt,
\newblock ``Cavity optomechanics'',
\newblock {\em Rev.~Mod.~Phys.} 86, 1391--1452 (2014).

\bibitem{Rossi:2018aa}
M.~Rossi, D.~Mason, J.~Chen, Y.~Tsaturyan, and A.~Schliesser,
\newblock ``Measurement-based quantum control of mechanical motion'',
\newblock {\em Nature} 563, 53--58 (2018).

\bibitem{Liu:2025aa}
Y.~Liu, H.~Sun, Q.~Liu, H.~Wu, M.~A.~Sillanp{\"a}{\"a}, and T.~Li,
\newblock ``Degeneracy-breaking and long-lived multimode microwave electromechanical systems enabled by cubic silicon-carbide membrane crystals'',
\newblock {\em Nat.~Commun.} 16, 1207 (2025).

\bibitem{10.1063/1.2868810}
J.~P.~Zendri, M.~Bignotto, M.~Bonaldi, M.~Cerdonio, L.~Conti, L.~Ferrario, N.~Liguori, A.~Maraner, E.~Serra, and L.~Taffarello,
\newblock ``Loss budget of a setup for measuring mechanical dissipations of silicon wafers between 300 and 4k'',
\newblock {\em Rev.~Sci.~Instrum.} 79, 033901 (2008).

\bibitem{10.1063/5.0055954}
E.~Serra, A.~Borrielli, F.~Marin, F.~Marino, N.~Malossi, B.~Morana, P.~Piergentili, G.~A.~Prodi, P.~M.~Sarro, P.~Vezio, D.~Vitali, and M.~Bonaldi,
\newblock ``Silicon-nitride nanosensors toward room temperature quantum optomechanics'',
\newblock {\em J.~Appl.~Phys.} 130, 064503 (2021).

\bibitem{10.1063/1.4938747}
M.~Yuan, M.~A.~Cohen, and G.~A.~Steele,
\newblock ``Silicon nitride membrane resonators at millikelvin temperatures with quality factors exceeding 108'',
\newblock {\em Appl.~Phys.~Lett.} 107, 263501 (2015).

\end{thebibliography}

\end{document}